\documentclass[prl,twocolumn,amssymb,showpacs]{revtex4}

\usepackage{graphicx}
\usepackage{color}
\usepackage{epsfig}
\usepackage{latexsym}
\usepackage{bm}
\usepackage{ulem}
\usepackage{color}

\usepackage{dcolumn}
\usepackage{amsmath}    
\usepackage{amssymb}
\usepackage{bm} 
\usepackage{hyperref}
\usepackage{latexsym}
\usepackage{color}
\usepackage{subfigure}
\def\beq{\begin{equation}}
\def\eeq{\end{equation}}

\def\Q{\mbox{\sffamily\bfseries Q}}

\def\lsim{\:\raisebox{-0.5ex}{$\stackrel{\textstyle<}{\sim}$}\:}
\def\gsim{\:\raisebox{-0.5ex}{$\stackrel{\textstyle>}{\sim}$}\:}
\def\beq{\begin{equation}}                           
\def\eeq{\end{equation}}                           
\def\bea{\begin{eqnarray}}                           
\def\eea{\end{eqnarray}}        

\begin{document}


\title{Density Induced Phases in Active Nematic}




\author{Rakesh Das}
\author{Manoranjan Kumar}
\author{Shradha Mishra}
\email[]{shradha.mishra@bose.res.in}

\affiliation{S N Bose National Centre for Basic Sciences, Block JD, Sector III, Salt Lake, Kolkata 700098, India}


\begin{abstract}
We introduce a minimal model for a collection of self-propelled apolar active particles,
also called as `active nematic', on a two-dimensional substrate and study the order-disorder transition with the variation of density.
The particles interact with their 
neighbours within the framework of the Lebwohl-Lasher model and move asymmetrically, along their orientation, 
to unoccupied nearest neighbour lattice sites. At a density lower than the equilibrium isotropic-nematic 
transition density, the active nematic shows a first order transition  from the isotropic state to a banded 
state. The banded state extends over a range of density, and the scalar order parameter of the system shows a plateau 
like behaviour, similar to that of the magnetic systems. In the large density limit the active nematic shows 
a bistable behaviour between a homogeneous ordered state with global ordering and an inhomogeneous mixed state 
with local ordering. The study of the above phases with density variation is scant and gives significant insight 
of complex behaviours of many biological systems.

\end{abstract}


\maketitle


{\it Introduction} :--- Active systems are composed of {\it self-propelled} 
particles where each particle extracts energy from its surroundings and 
dissipates it through motion towards
a direction determined by its orientation. 
These kind of systems are ubiquitous in nature, ranging from very small scale systems inside the 
cell to larger scales  \cite{harada, nedelec, rauch, benjacob, animalgroups, helbing, feder, kuusela31, hubbard},
vibrated granular media \cite{vnarayan, kudrolli} etc.,
and have been studied extensively through experiments, theories and simulations  \cite{sriramrmp, tonertusr, rev}.
A collection of head-tail symmetric `apolar' active particles with an average mutual parallel alignment is said to 
be in a `nematic' state, whereas in an `isotropic' state particles remain randomly oriented. 
An active system where
fluid media do not play important role in emergence of ordered state, and thus the hydrodynamic interactions can be 
ignored, is called a `dry active system' \cite{kemkemer, vnarayan, animalgroups, serra, schaller, surrey}. 

Active nature of particles introduces a nonequilibrium coupling between density and orientation field,
as represented in terms of curvature coupling current in literature \cite{sradititoner, shradhanjop, sriramrmp}. 
Such coupling in active nematic induces unusual properties like large density fluctuation \cite{sradititoner, chateprl2006} 
and growth kinetics faster than $1/3$ as in usual conserved model \cite{shradhatrans2014}.
Recent studies of the active 
nematic  found a defect-ordered nematic  state  
 \cite{aparnaredner, shimanatcomm, yeomans} as opposed to
the equilibrium nematic for high particle densities. Recent experiment
on amolyiod flibrils \cite{ncommam} also found a phase with coexisting aligned and 
disordered fibril domains, similar to the defect-ordered 
state obtained in simulations. 
But few investigations have been done on the behaviours of the active nematic in various density
limits, especially at low densities.
Here we introduce a minimal 
model for two-dimensional active nematic and  compare  various ordering  phases of  active and equilibrium 
nematic in different density limits. 
The ordering in the system is characterised in terms of a scalar order parameter $S$ which is the positive eigen 
value of nematic order parameter $\Q$ \cite{pgdgenne} in two-dimensions. 
In the low density limit both active and equilibrium systems are in the isotropic (I) state with particles randomly oriented 
throughout the whole system (see Fig. \ref{fig:phase_snap}(b) - I), resulting in a small $S$.
The Phase diagram of the active nematic as a function of packing density $C$ (see Fig. \ref{fig:phase_snap}(a)) 
shows a jump in $S$ close to $C=0.37$, whereas in the equilibrium case $S$ goes continuously to larger values and 
an isotropic to nematic (I-N) transition occurs close to $C=0.58$. In the equilibrium nematic (EN) state particles 
remain homogeneously oriented in the system (see Fig. \ref{fig:phase_snap}(b) - EN). At $C=0.37$ 
the active system goes from the isotropic to a banded state (BS) 
where particles cluster and align in the perpendicular direction to the long axis of the band (see Fig. \ref{fig:phase_snap}(b) - BS-1). 
With increasing density more number of particles participate in band formation (see Fig. \ref{fig:phase_snap}(b) - BS-2) 
and $S$ follows a plateau over a range of density.
In the large density limit active system shows bistability between  
a homogeneous ordered (HO) (see Fig. \ref{fig:phase_snap}(b) - HO) 
and an inhomogeneous mixed (IM) or local ordered state (see Fig. \ref{fig:phase_snap}(b) - IM). 
This IM state is very similar to defect-ordered nematic state in ref. \cite{aparnaredner, shimanatcomm, yeomans}.


\begin{figure*}[t]
  \centering
  \subfigure[]{{\includegraphics[height=5.95cm]{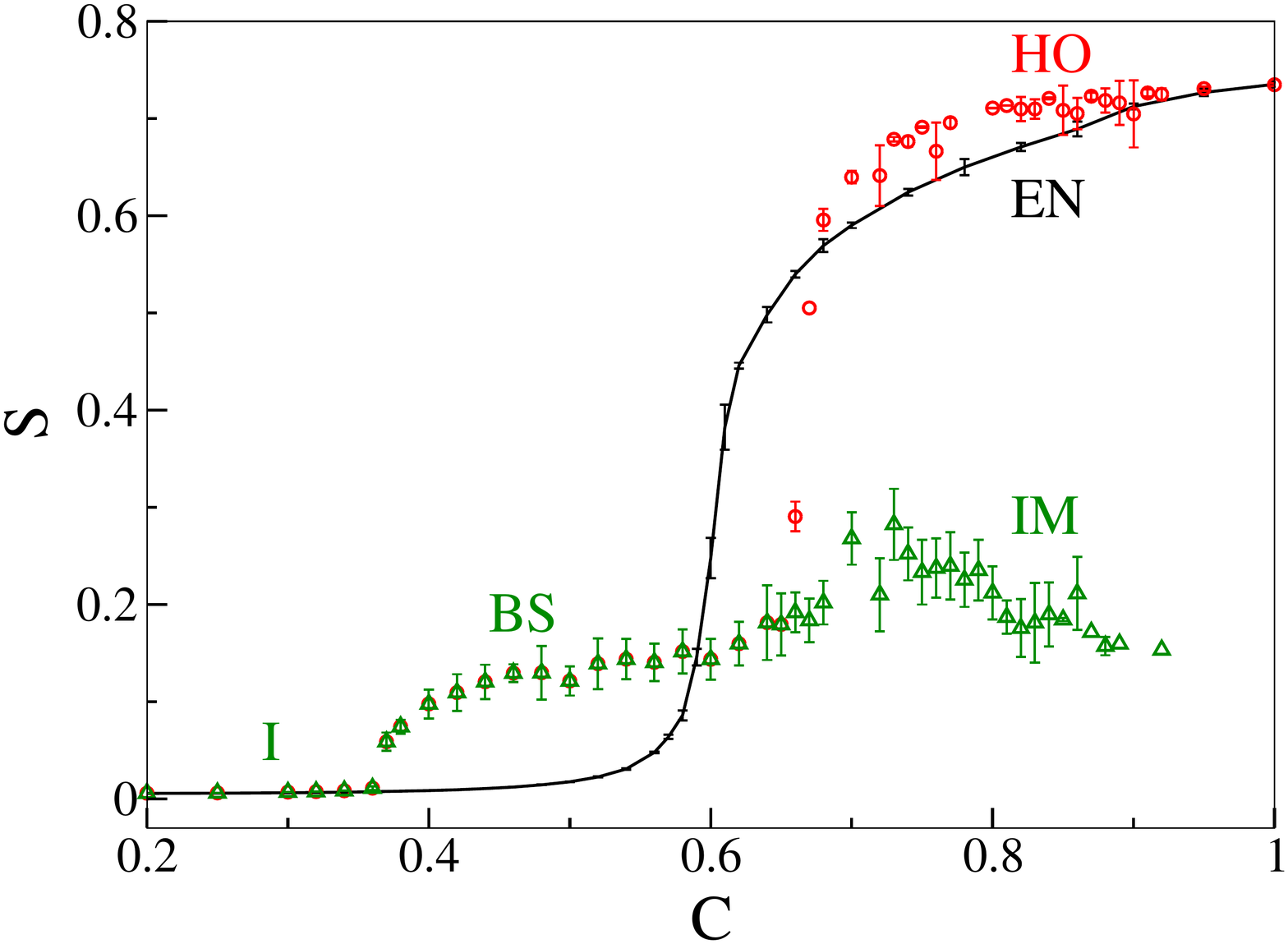}}}
  \subfigure[]{{\includegraphics[scale=0.35]{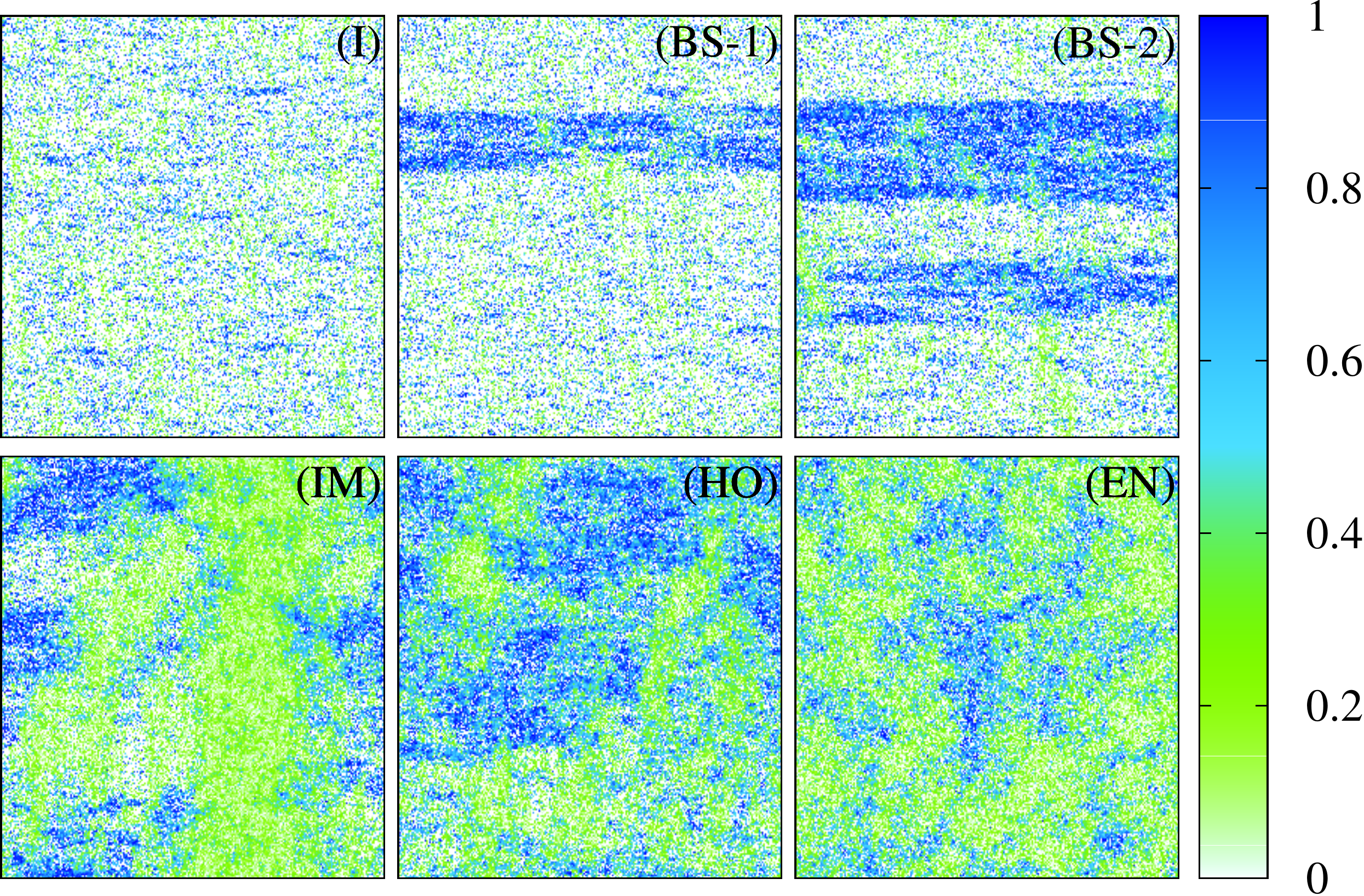}}}
\caption{(Color online) (a) Plot of scalar order parameter $S$ vs. packing density $C$ for
 active (circles and triangles) and equilibrium (continuous line) 
nematic for system size $512$ $\times$ $512$. 
Equilibrium system goes smoothly from isotropic (I)
to nematic (EN) state. 
Active system goes from isotropic (I) to banded state (BS)
 (small jump in $S$)  followed by
either an inhomogeneous mixed (IM) (triangles) 
or a homogeneous ordered (HO) (circles) state. 
(b) Snapshots of particle inclination towards 
the horizontal direction. 
Color bar ranging from zero to one
indicates  parallel  and  perpendicular inclinations
respectively towards  the horizontal direction. 
White regions signify unoccupied sites.
(I) is isotropic state at low density ($C=0.36$), 
(BS-1) ($C=0.38$) and (BS-2) ($C=0.56$) are two 
banded state configurations, (IM) is inhomogeneous mixed, 
 (HO) is homogeneous ordered  and   
(EN) is equilibrium nematic state at high density ($C=0.76$).}
  \label{fig:phase_snap}
\end{figure*}


{\it Model} :--- We consider a two dimensional square lattice. At each vertex `$i$'  we define an occupation
variable $n_i$, which can take values $1$ (occupied) or $0$ (unoccupied), and an orientation variable $\theta_i$, which 
lies between $0$ and $\pi$ because of the apolar nature of the particles. Each particle interacts with its nearest neighbours 
through  modified Lebwohl - Lasher  Hamiltonian  \cite{llasher}
\begin{equation}
\mathcal{H} = -\epsilon \sum_{<ij>}n_i n_j \cos2(\theta_i-\theta_j)
\label{eqll}
\end{equation}
where $\epsilon$ is the interaction strength between two neighbouring particles. This model is analogous to the diluted XY-model 
with nonmagnetic impurities \cite{dilutedxymodel}, where impurities and spins are analogous to vacancies and particles 
respectively in the present model.

Orientation evolves through Monte - Carlo (MC) updates \cite{mcbinder} following the Hamiltonian in Eq. \ref{eqll}.
Unlike the diluted XY-model, particles also move on the lattice. Depending on the type of movement we define two kinds of models.
(i)  `Equilibrium model' (EM) - a particle can diffuse to any unoccupied nearest-neighbouring site, and therefore
satisfies the detailed balance condition. (ii) `Active model' (AM) - a particle can move to only those
unoccupied nearest-neighbouring sites which are in the direction that makes the least inclination with the particle orientation.
Details of the model and particle movement are shown in Supplemental Material \cite{SM} section I.
The asymmetric move of the active particles does 
not staisfy the detailed balance condition and arises in general
because of the self-propelled nature of the  particles in many biological \cite{kemkemer, paxton} and granular systems \cite{vnarayan}. 
These moves produce an active curvature coupling current  in coarse-grained hydrodynamic 
equations of motion \cite{shradhanjop, sradititoner}.

{\it Numerical details} :---
We consider a collection of $N$ particles with random orientation $\theta_i \in [0,\pi]$, homogeneously distributed
on a $L \times L$ square lattice ($L=150, 256, 512$) with periodic boundary condition. The packing density of the system
is $C=N/(L \times L)$.
We choose a particle randomly and move it to an unoccupied neighbouring site, followed by an orientation updation through MC.
We use $10^6$ MC steps to evolve the system to 
its steady state and all the results have been obtained by averaging over next $2 \times 10^6$ MC steps. 
Twenty four realizations have been used for better statistics.

We calculate the scalar order parameter
\begin{equation}
S=\sqrt{(\frac{1}{N}\sum_i n_i \cos(2 \theta_i))^2+(\frac{1}{N}\sum_i n_i \sin(2 \theta_i))^2}
\label{eqops}
\end{equation}
which is small in the isotropic state and close to $1$ in the ordered  state. 
First we calculate $S$ for EM as a function of inverse temperature 
$\beta= 1 / k_BT$ for different densities.
As  shown in Supplemental Material \cite{SM} section II, the critical temperature $T_c$ is 
approximated as $T_c(S=0.4)$.
Critical temperature  $T_c(C)$ decreases with the lowering of the packing density $C$
, similar trend   is found  in the study of diluted XY-model \cite{dilutedxymodel} for varying nonmagnetic site density. 
In rest of our calculations temperature is kept fixed at $\beta\epsilon = 2.0$ 
and packing density $C$ is varied from small values to complete filling $C=1.0$.


{\it Phase diagram} :---
At low densities, $C<0.37$, the active system is in the isotropic state where the particles with random orientation 
remain homogeneously distributed throughout the system, and therefore $S$ holds vanishingly small values.
The jump occurs in $S$  at $C=0.37$. For $C \geq 0.37$  particles cluster in, and both ordered state with high 
local density and disordered state with low local density coexist (see Fig. \ref{fig:phase_snap}(b) - BS-1).
 Mean alignment inside the band is perpendicular to the long axis of the band. 
In the moderate density limits, band formation in more favourable than lane formation 
(mean alignment parallel to the long axis of the structure) because the number of particles that can have translational 
motion is much larger in the banded state, and therefore entropy favours band formation.
Similar mechanism create the 
bending instability close to order-disorder transition in the active  polar systems \cite{chate, shradhapre}.
The above transition from I state to BS occurs at density lower than the corresponding 
equilibrium I-N transition density $C \simeq 0.58$ (see Fig. \ref{fig:phase_snap}(a)). 

As we  further increase $C$, unlike the equilibrium 
system where $S$ increases monotonically with $C$,  
the active system  shows very small change in $S$ for a range of density.
This plateau like appearance of $S$ with variation in $C$ is very similar to plateau
phase in magnetization versus field curve of magnetic systems \cite{plateauphase}.
If an energy gap exists between two consecutive magnetic states, a finite field is required
for the magnetic system to go from the lower to the higher state. So until that finite field 
is applied, the increasing field keeps the system magnetization to be unchanged, and the system
is called to be in the plateau phase. With increasing packing density in the plateau regime of 
the active nematic more particles participate in band formation (see Fig. \ref{fig:phase_snap}(b) - BS-1 and BS-2). 
On further increment of density,  close to equilibrium I-N transition $C \simeq 0.58$, 
transverse fluctuations  lead the system to a mixed state \cite{shradhanjop, shimaprl2011}.

In the large $C$ limit active system shows a bistable behaviour with two distinct steady states;
first, a state where $S$ is large and real space configuration is  `homogeneous ordered' (HO),
and the second, an `inhomogeneous mixed' (IM) state where $S$ realizes some moderate  values. 
In the HO state though the particle orientation is homogeneous, large density inhomogeneity exists 
in the system (see Fig. \ref{fig:phase_snap}(b) - HO). IM  state is a local ordered state with many 
aligned clusters of high particle density. The system consists of many such aligned 
clusters of high density separated from low density disordered regions 
and mean alignment in each cluster is in different directions 
(see Fig. \ref{fig:phase_snap}(b) - IM). IM state is similar to the defect-ordered state
recently found in the study of ref. \cite{shimanatcomm, aparnaredner, yeomans},
with large number of $\pm 1/2$ defects in high density active nematic.

{\it Renormalised mean field study for small $S$}:--- 
We also calculate the jump  in the scalar order parameter $S$
and the shift  in the transition density using  the Renormalised mean field (RMF) method of
the coupled coarse-grained hydrodynamic equations of motion for the number density
$\rho({\bf r}, t) = \sum_{\alpha}\delta({\bf r}-{\bf R}_\alpha(t))$ and the  order
parameter $ {\bf w}_{i j}({\bf r}, t)  = \rho({\bf r}, t) \Q({\bf r}, t) = \sum_{\alpha} ({\bf m}_{i \alpha} {\bf m}_{j \alpha} - \frac{1}{2}\delta_{i j}) 
\delta({\bf r}-{\bf R}_{\alpha}(t))$
for active nematic \cite{shradhanjop, sradititoner}.
\begin{equation}
\partial_{t}\rho=a_{0}\nabla_{i}\nabla_{j}{\bf w}_{ij}+D_{\rho}\nabla^{2}\rho
\label{eqdensity}
\end{equation}
and
\begin{align}
\partial_{t}{\bf w}_{ij} & =\left\{ \alpha_{1}\left(\rho\right)-\alpha_{2}\left(\mathbf{w}:\mathbf{w}\right)\right\} {\bf w}_{ij} \notag \\
 & + \beta\left(\nabla_{i}\nabla_{j}-\frac{1}{2}\delta_{ij}\nabla^{2}\right)\rho+D_{\bf w}\nabla^{2}{\bf w}_{ij}
\label{eqop}
\end{align}
where, ${\bf m}_{\alpha} = (\cos(\theta_{\alpha}), \sin(\theta_{\alpha}))$
is the unit vector along the orientation $\theta_{\alpha}$ and  ${\bf R}_{\alpha}(t)$ is 
the position of particle $\alpha$.
We can obtain the number density $\rho$ by coarse-graining $C$ over small subvolume.
  Eqs. \ref{eqdensity} and \ref{eqop} can be obtained
either from symmetry arguments as in ref. 
\cite{sradititoner} or from  microscopic rule
based model \cite{shradhanjop}.
Density equation  \ref{eqdensity} is a continuity
equation $\partial_{t}\rho = -\nabla \cdot {\bf J}$,
because the total number of particles
is conserved. Current $J_i = -a_0 \nabla_j {\bf w}_{ij} - D_{\rho} \nabla_i \rho$,
where the first term consists of two parts, an active curvature coupling 
current ${\bf J}_a \propto a_0 {\rho \nabla_j \Q_{ij}}$ and anisotropic 
diffusion ${\bf J}_{p1} \propto \Q_{ij}\nabla_i \rho$, which can also
be present in the equilibrium model.
The second term in density equation is an isotropic diffusion  ${\bf J}_{p2} \propto \nabla \rho$ term.
First two terms in the order parameter equation ${\bf w}_{ij}$ is the alignment
term. We choose $\alpha_1(\rho) = (\frac{\rho}{\rho_{IN}}-1)$ as
a function of density which changes sign at some critical
density $\rho_{IN}$. Third term is
coupling to density and last term is diffusion in order
parameter and written for equal elastic constant approximation
for two-dimensional nematic. 

\begin{figure}[t]
 \includegraphics[width=\linewidth]{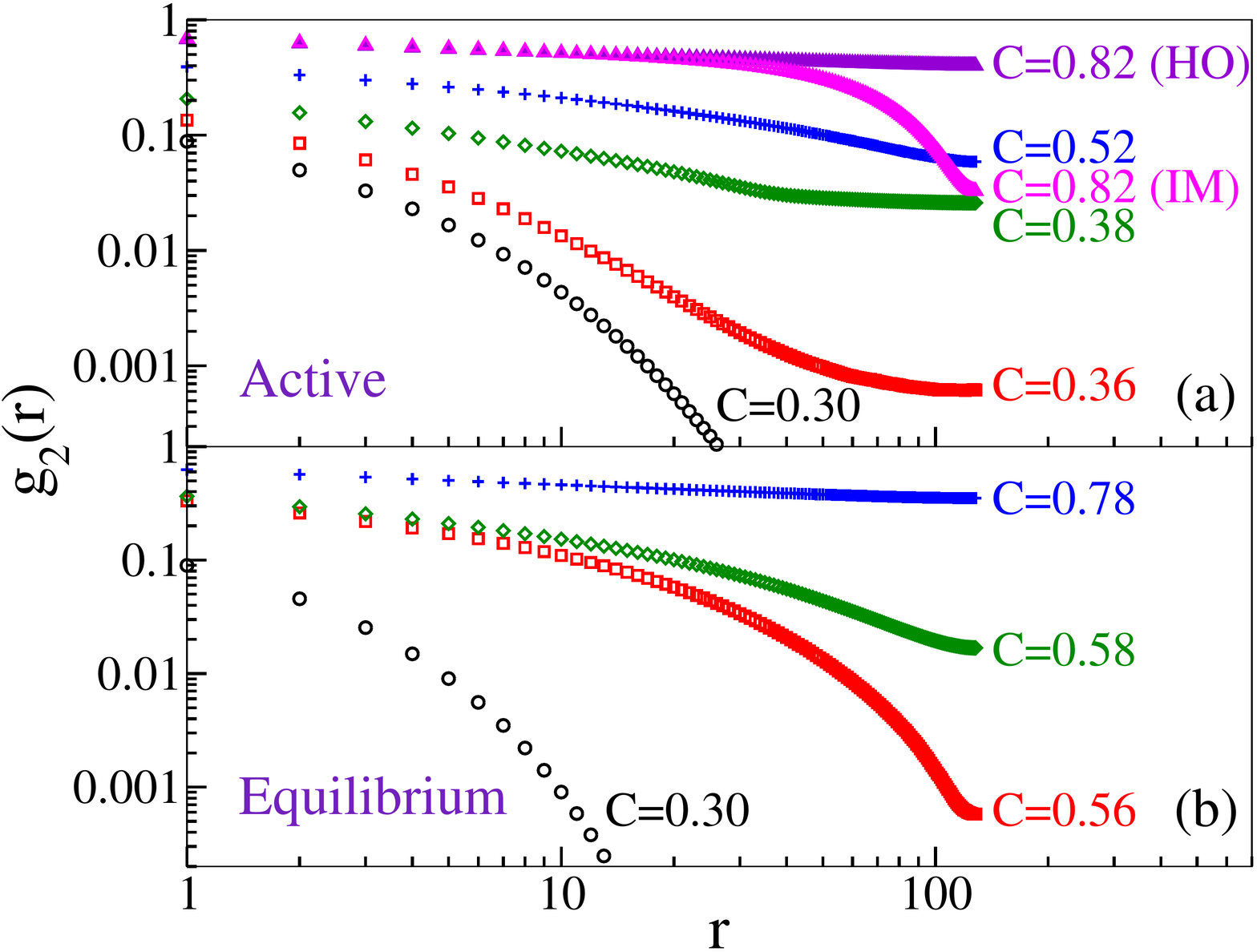}
 \caption{(Color online) $g_2(r)$ vs. $r$ on log-log scale
 at different densities.
(a) Active nematic: ($\bigcirc$) and ($\square$) at low densities
$g_2(r)$ decays exponentially,  ($\diamond$) and
($+$) at intermediate density $g_2(r)$ decays algebraically,
($\bigtriangleup$)  homogeneous ordered (algebraic decay),
and inhomogeneous mixed (abrupt decay)  at high density.
(b) Equilibrium nematic ($\bigcirc$) and ($\square$)
exponential decay of $g_2(r)$  at low densities and
($\diamond$) and ($+$) algebraic decay   of
 $g_2(r)$ at high densities.}
 \label{fig:correlation}
\end{figure}

A homogeneous steady state solution of Eqs. \ref{eqdensity}
 and \ref{eqop} gives a mean field transition
from isotropic to nematic state at density
$\rho_{IN}$ where $\alpha_1(\rho)$ changes sign.
Using RMF method we calculate the effective free
energy $f_{eff}(S)$
 close to order-disorder transition when $S$ is small. We consider density fluctuations 
 $\delta \rho$ and neglect order parameter fluctuations.
  The effective free energy
\begin{equation}
f_{eff}\left(S\right)=-\frac{b_{2}}{2}S^{2}-\frac{b_{3}}{3}S^{3}+\frac{b_{4}}{4}S^{4}
\label{eqfenergy}
\end{equation}
where $b_{2}=\alpha_{1}(\rho)+\alpha_{1}^{\prime}(\rho)c$, where
$c$ is a constant.
$\alpha_1^{\prime}(\rho) = {\partial \alpha_1/\partial \rho|}_{\rho_0}$,
$b_{3}=\frac{a_{0}\alpha_{1}^{\prime}(\rho)}{2D_{\rho}}$
and $b_{4}=\frac{1}{2}\alpha_{2}$.
Both $b_3$ and $b_4$ are positive. 
A detail calculation for $f_{eff}$ is shown in Supplemental Material \cite{SM} section III.
The density flcutuations introduce a new cubic order term  protortional to the activity strength $a_0$, in  the free energy $f_{eff} (S)$.
The presence of such term produces a jump $\Delta S = S_{c}=\frac{2b_{3}}{3b_{4}} $  at a lower density 
$\rho_c = \rho_{IN}(1-\frac{2b_3^2}{9b_4} ) $. This type of jump and shift in transition because of flcutuations
are also called as fluctuation dominated first order phase transition in statistical mechanics \cite{coleman} and widely
studied in many systems \cite{fdfopt}.
The jump in  $S$ and the shift in $\rho_c$ is proportional to the activity parameter
$a_0$ and for $a_0=0$ we recover the equilibrium transition.

{\it Two-point orientation correlation function} :---
To further characterise the system we also calculate
  the two-point orientation correlation
$g_2(r) = <\sum_i n_i n_{i+r} \cos[2\left(\theta_i-\theta_{i+r}\right)] / \sum_i n_i n_i > $ at different packing densities,
where $< . >$ signifies an average over many realisations.
In Fig. \ref{fig:correlation} we plot $g_2(r)$ vs. $r$ on log-log scale, for $C=0.30$, $0.36$, $0.38$, $0.52$ and $0.82$ for
active model and $C=0.30$, $0.56$, $0.58$ and $0.78$ for equilibrium model.
For very small packing density $C<0.37$, $g_2(r)$ decays exponentially in the active systems.
Therefore the system is in short-range-ordered (SRO) isotropic state.
At $C=0.38$, $g_2(r)$ decays following the power law $g_2(r) \simeq 1/r^{\eta(C)}$ and
therefore the system is in a quasi-long-range-ordered (QLRO) state.
At high packing densities correlation functions  confirm the bistability in the active systems.
For $C=0.82$ (see Fig. \ref{fig:correlation}(a)) $g_2(r)$ shows power law decay in HO state,
whereas in IM state $g_2(r)$ decays abruptly after a distance $r$. The abrupt change in $g_2(r)$ at a certain distance indicates
the presence of local ordered clusters.
In contrast, the equilibrium systems show a transition from SRO isotropic state at low density $C\lsim0.56$
to QLRO nematic state at $C\gsim0.58$, similar to Berezinskii - Kosterlitz - Thouless (BKT) transition \cite{Berezinskii, KT}
in the diluted XY-model \cite{dilutedxymodel}.

{\it Orientation distribution} :---
We also compare  the steady state properties of active and equilibrium models
in the high density limit. First we calculate the steady state (static) 
orientation distribution $P(\theta)$
from one  snapshot of particle orientation. Both active HO and
equilibrium nematic show a  Gaussian distribution of orientation (see Fig. \ref{fig:distribution}(a)).
Hence in HO state orientation fluctuations of particles are of same kind as in equilibrium model
and  the system is in QLRO state. 
Distribution $P(\theta)$ in the IM state is very broad and spanning the 
whole range of orientation. Hence the system has many
 local ordered regions with different orientations.

We also calculate the time averaged distribution of mean orientation of all the particles
 $P(\theta_m)$ in active HO and equilibrium nematic states. $P(\theta_m)$ is 
calculated from mean of all particle orientaions averaging over a long time (from $10^6$ to $3 \times 10^6$)
 in the steady state. $P(\theta_m)$ for HO is narrow in comparison to the broad distribution 
for equilibrium model (see Fig. \ref{fig:distribution}(b)). Narrow distribution of $P(\theta_m)$
implies that orientation autocorrelation in active system decay slowly as comapared to the 
corresponding equilibrium model which is in agreement with  the slow decay of velocity autocorrelation
in bacterial suspension \cite{wulib}.

\begin{figure}[t]
 \includegraphics[width=\linewidth]{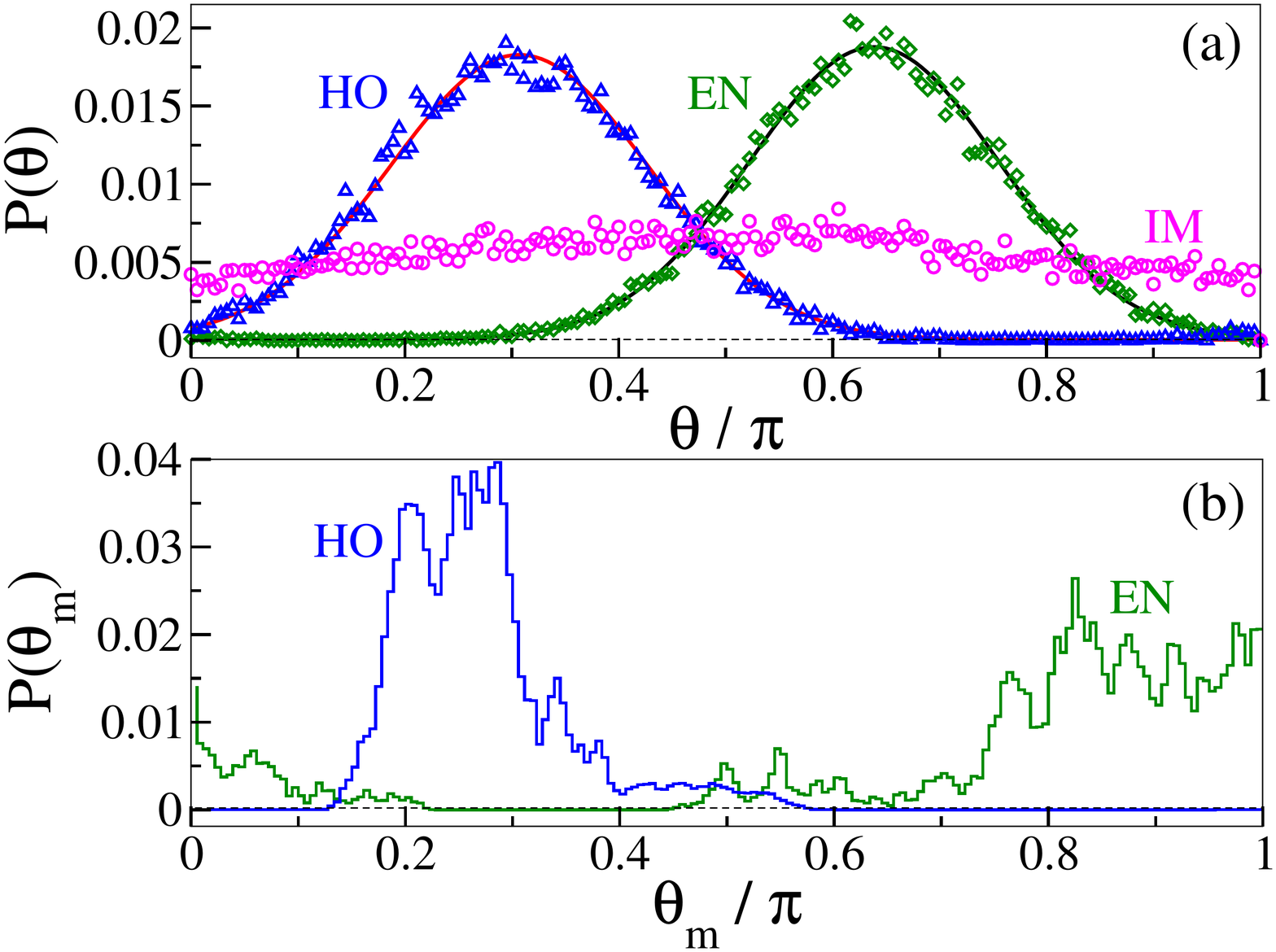}
 \caption{(Color online) (a) Steady state orientation distribution $P(\theta)$
 of particles  for
HO, IM (active) and  EN (equilibrium)  states at high density. 
Lines are fit to Gaussian distribution for
both HO and EN states. IM state shows very broad distribution 
of $P(\theta)$. (b) Plot of mean orientation distribution $P(\theta_m)$
averaged over a long
time  in the steady state for HO (active) and EN (equilibrium) states. 
$P(\theta_m)$ is very
broad for EN in comparison to HO.}
 \label{fig:distribution}
\end{figure}

{\it Summary} :--- 
In this letter we have introduced a minimal model for active nematic and  
found three distinct phases with the variation in density.
At low densities the active nematic is in disordered isotropic state with
very small correlation between the particles.
With increasing density active nematic undergoes a fluctuation induced first order
phase transition from the isotropic to the banded state where large number of particles participate in band formation.
Large density fluctuations in the active systems change the nature of the transition and shift the 
transition  density to smaller value as compared to the equilibrium isotropic nematic transition.
 At large densities  equilibrium nematic is in QLRO nematic state,
whereas active nematic goes from the banded state to either the homogeneous ordered (high $S$)  
or the inhomogeneous mixed (moderate  $S$) state. This inhomogeneous mixed state is similar to the
phase with coexisting aligned and disordered fibril domains found in recent experiment \cite{ncommam}.
Experiments on thin layer  of agitated  granular rods, elongated living cells, bacterial colony of 
apolar {\it B. subtilis} etc. at different
densities can realize the different phases we found here.
In the present model we have frozen the  motion of the active particles in the transverse direction, 
{\it i.e.} the activity strength is kept large.
It will be interesting to see the evolution of different phases with the particles having a small
probability to move in transverse directions as well.

\begin{acknowledgments}
S. M. acknowledges Thomas Niedermayer for useful discussions. S. M. and M. K. acknowledges financial support from the Department of Science and Technology, India, under INSPIRE award 2012 and Ramanujan Fellowship respectively.
\end{acknowledgments}



\newpage

\begin{titlepage}\centering
{\bf \Large Supplemental Material}
\end{titlepage}
\onecolumngrid

\section{I. Model figure}

\begin{figure*}[ht]
  \includegraphics[width=0.8\linewidth]{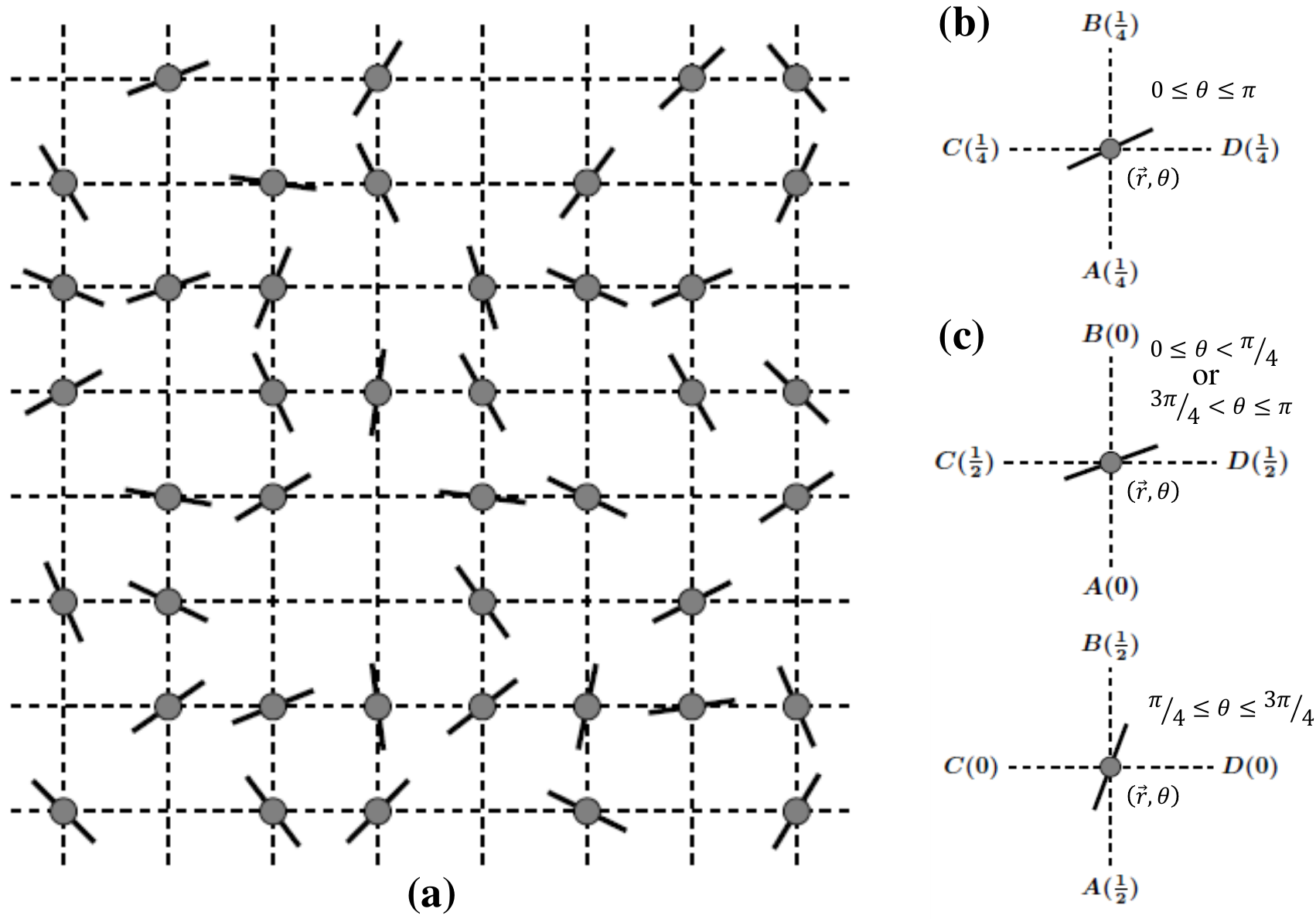}
  \caption{(a) Two dimensional square lattice with occupied ($n=1$) or unoccupied ($n=0$)
sites. Filled circles signify the occupancy of respective sites. Inclination of
the rods towards the horizontal direction show respective particle orientation $\theta_i\in\left[0,\pi\right]$.
(b) Equilibrium move : particle can move to any of four neighbouring sites with equal probability $1/4$,
 (c) Active move: particle can move to either of its two
neighbouring sites with probability $1/2$, if unoccupied, in the direction it  is more inclined to.}
  \label{fig:model}
\end{figure*}

\newpage

\section{II. Estimate of critical temprature $T_c(C)$ in equilibrium model} 

\begin{figure}[ht]
  \includegraphics[width=0.5\linewidth]{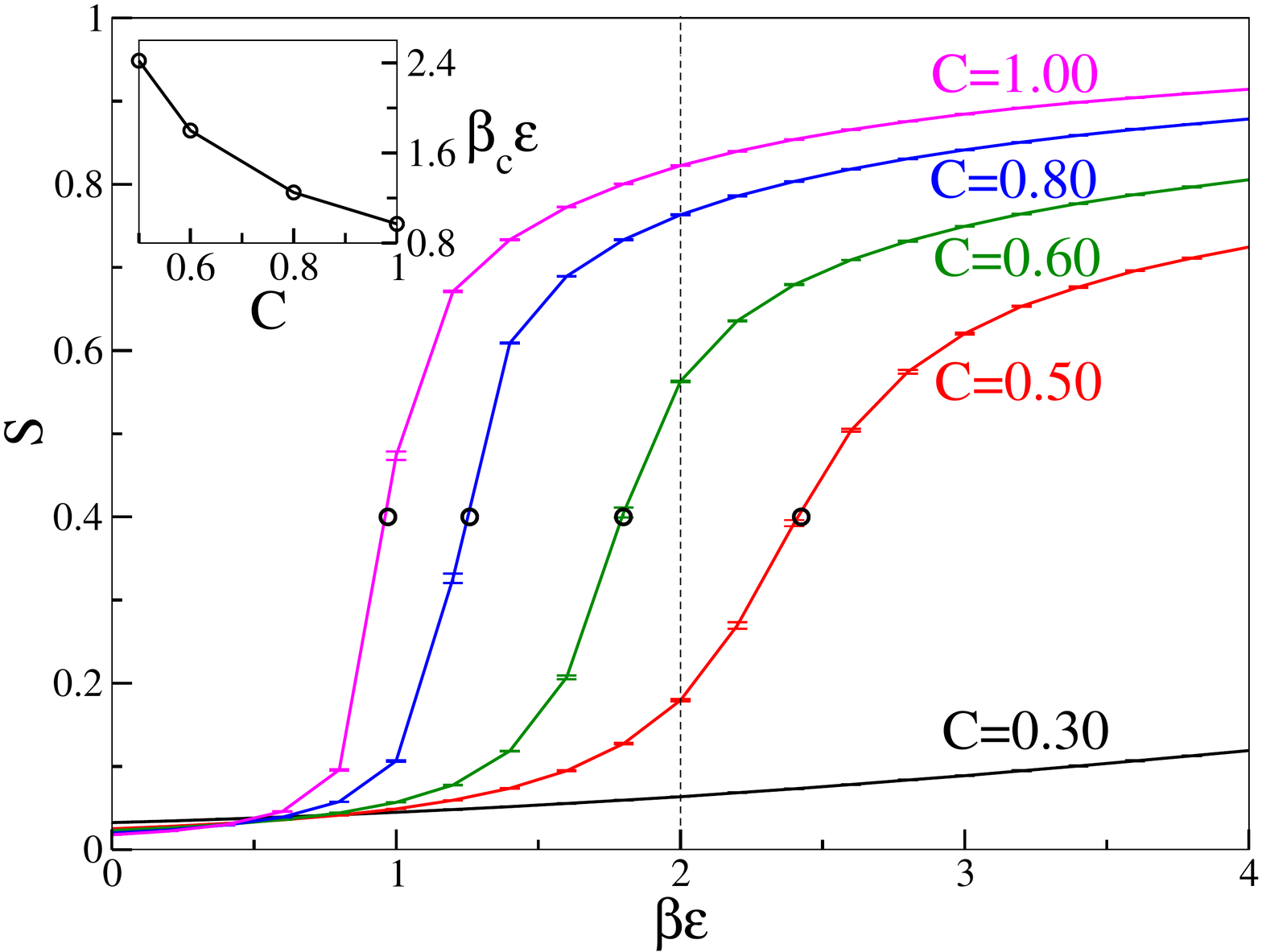}
\caption{Plot of $S$ vs. inverse temperature $\beta\epsilon$ for different densities $C$ for equilibrium model. System goes from
isotropic (small $S$) to nematic (large $S$) state. Vertical dotted line shows the variation in $S$ for fixed $\beta\epsilon=2.0$ at different densities. Crtical temperature is approximated as $T_c(S=0.4)$. Inset: change in
$T_c$ as a function of density $C$.}
\label{figtemp}
\end{figure}


\section{III. Renormalised mean field (RMF) study of active nematic for small scalar order parameter $S$}

In this section we will write an effective renormalised mean field free energy 
 for scalar  order parameter $S$ for small $S$. We keep the density fluctuations 
 and ignore the order parameter fluctuations in the coupled
hydrodynamic equations of motion for active nematic. 
Density fluctuation produces a  cubic  order term in the effective
free energy for scalar order parameter $S$ and such term produces a jump 
in $S$ at a new transition density $\rho_c$ lower than 
equilibrium I-N transition point $\rho_{IN}$. 
Shift in transition density and jump $\Delta S$ is directly 
proportional to the activity
parameter $a_0$ and we recover equilibrium limit for zero $a_0$. \\
We write coupled hydrodynamic equations of motion for density $\rho$ and 
 order parameter ${\bf w} = \rho \Q$ 
where nematic order parameter \cite{pgdgenne}
\begin{eqnarray}			
\Q({\bf r}, t)=S\left(\begin{array} {c c}
\cos 2\theta ({\bf r}, t) & \sin 2\theta ({\bf r}, t) \\
\sin 2\theta ({\bf r}, t) & -\cos 2\theta ({\bf r}, t)
\end{array} \right)
\label{Q_tensor}
\end{eqnarray}
$\theta$ being the coarse grained 
angle at position ${\bf r}$ and time $t$.
Density equation
\begin{equation}
\partial_{t}\rho=a_{0}\nabla_{i}\nabla_{j}{\bf w}_{ij}+D_{\rho}\nabla^{2}\rho
\label{eqa1}
\end{equation}
and order parameter equation ${\bf w}$
\begin{equation}
\partial_{t}{\bf w}_{ij}=\left\{ \alpha_{1}\left(\rho\right)-\alpha_{2}\left(\bf{w}:\bf{w}\right)\right\} {\bf w}_{ij}+\beta
\left(\nabla_{i}\nabla_{j}-\frac{1}{2}\delta_{ij}\nabla^{2}\right)\rho+D_{{\bf w}}\nabla^{2}{\bf w}_{ij}
\label{eqa2}
\end{equation}
Density Eq. \ref{eqa1} is a continuity equation $\partial \rho/\partial t = - \nabla  \cdot {\bf J}$,
where ${\bf J}$ has two parts, active and diffusive. Details of these two currents are given
in the main text.  $a_0$ is the activity parameter, present  because of self-propelled nature
of the particles,  $\beta$ is the coupling of density in ${\bf w}$ equation. 
$D_{\rho}$ and $D_{\bf w}$ are the diffusion coefficients  in density and order parameter equations
respectively, $\alpha_1(\rho)$ and $\alpha_2$ are the alignment
terms and ingeneral depends on the model parameters. For  metric distance
interacting models \cite{njopshradha} $\alpha_1(\rho)$ is a 
function of density and changes sign at some 
critical density. We choose $\alpha_1(\rho)= \frac{\rho}{\rho_{IN}}-1$ and 
$\alpha_2 = 1$. Steady state solution of homogeneous Eq. \ref{eqa1} is
$\rho=\rho_0$, we add small perturbation to mean density $\rho= \rho_0 + \delta \rho$.
In the staedy state  density fluctuation $\delta \rho$  can be obtained from Eq. \ref{eqa1},
\begin{align}
& a_{0}\nabla_{i}\nabla_{j}{\bf w}_{ij}+D_{\rho}\nabla^{2}\delta \rho = 0 \notag \\
& \Rightarrow 
 a_{0}\nabla_{j}{\bf w}_{ij}+D_{\rho}\nabla_{i}\delta \rho=constant=\mathbf{c_{1}} 
\label{eqa3}
\end{align}
where  ${\bf w}_{11} = -{\bf w}_{22} =  \frac{S}{2} \cos(2 \theta)$ and 
${\bf w}_{12} = {\bf w}_{21} =  \frac{S}{2} \sin(2 \theta)$ and keep the lowest order
 terms in $S$ and $\theta$
\begin{equation}
\partial_x \delta \rho = -\frac{a_0}{D_{\rho}} \partial_x S \rightarrow \delta \rho(x)=-\frac{ a_0}{D_{\rho}} S + c
\label{eqa4}
\end{equation}
and 
\begin{equation}
 \partial_y \delta \rho = \frac{a_0}{D_{\rho}} \partial_y S \rightarrow \delta \rho(y)=\frac{a_0}{D_{\rho}} S + c_1
\label{eqa5}
\end{equation}
Here we assume nematic is aligned along one direction and
there is variation only along the perpendicular direction.
Hence we can choose either of equations \ref{eqa4} or \ref{eqa5}.
Two constants $c$ and $c_1$ are the value of density where 
nematic order parameter is zero. 

We use Eq. \ref{eqa4} and substitute the solution for density in equation
for ${\bf w}_{ij}$ and obtain an effective equation for $S$.
\begin{equation}
\partial_{t}S=\left\{ \alpha_{1}\left(\rho\right)-\frac{1}{2}\alpha_{2}S^{2}\right\} S + \mathcal{O}(\nabla^2 S) + \mathcal{O}(\nabla^2 \rho)
\label{eqa6}
\end{equation}
We neglect all the derivative terms and keep only  polynomial in $S$,
i.e. we neglect higher order fluctuations.
We can do taylor expansion of $\alpha_1(\rho)$ 
about mean density $\rho_0$,  
$\alpha_1(\rho)=\alpha_1(\rho_{0}+\delta \rho)=\alpha_1(\rho_{0}) + \alpha_1^{\prime} \delta \rho$,
where $\alpha_1^{\prime} = \frac{\partial \alpha_1}{\partial \rho}|_{\rho_{0}}$.
  Substitute the expression 
for $\delta \rho$ from Eq. \ref{eqa4} hence 
\begin{equation}
\partial_{t}S=\left\{ \alpha_{1}\left(\rho_{0}\right)+\alpha_1^{\prime} \delta \rho-\frac{1}{2}\alpha_{2}S^{2}\right\} S
\label{eqa7}
\end{equation}
We can write an effective free energy for $S$
\begin{equation}
\partial_t {S} = -\frac{\delta f_{eff}(S)}{\delta S}
\label{eqa8}
\end{equation}
hence
\begin{equation}
-\frac{\delta f_{eff}}{\delta S}  =S\left\{{ \alpha_{1}\left(\rho_{0}\right)+\alpha} \left(\rho_{0}\right)\left(\frac{a_{0}}{2D_{\rho}}S+c_{1}\right)-\frac{1}{2}\alpha_{2}S^{2}\right\}  
\end{equation}
Therefore
\begin{equation}
f_{eff}\left(S\right)=-\frac{b_{2}}{2}S^{2}-\frac{b_{3}}{3}S^{3}+\frac{b_{4}}{4}S^{4}
\label{eqa9}
\end{equation}
where $b_{2}=\alpha_{1}\left(\rho_{0}\right)+\alpha_1^{\prime} c$,
$b_{3}=\frac{a_{0}\alpha_1^{\prime} }{2D_{\rho}}$
and $b_{4}=\frac{1}{2}\alpha_{2}$ and $c$ is a conatant.
 Hence fluctuation in density 
introduces a cubic order term in effective free energy $f_{eff}(S)$. 
Effective free energy in Eq. \ref{eqa9} is similar to Landau free energy with cubic order
term \cite{chaiklub}. We calculate jump $\Delta S$ and new critical density 
 from coexistence condition
for free energy. Steady state solutions of order parameter 
($S=0$ for isotropic and $S \neq 0$ for ordered state) are given by 
\begin{equation}
\frac{\delta f_{eff}}{\delta S}=\left(-b_{2}-b_{3}S+b_{4}S^{2}\right)S=0
\end{equation}
Non-zero $S$ is given by $-b_{2}-b_{3}S_c+b_{4}S_c^{2}=0$. Coexistence condition implies
\begin{equation}
f_{eff}(S_c)=\left(-\frac{b_{2}}{2}-\frac{b_{3}}{3}S_c+\frac{b_{4}}{4}S_c^{2}\right)S_c^{2}=f_{eff}(S=0)=0
\end{equation}
hence we get the solution
\begin{equation}
S_{c}=-\frac{3b_{2}}{b_{3}}
\end{equation}
and
\begin{equation}
b_{2}^{c}=-\frac{2b_{3}^{2}}{9b_{4}}
\end{equation}
Hence the jump at new critical point is $\Delta S = \frac{2 b_3}{3 b_4}$.
Since $b_4 >0$ and hence $b_2^c <0$, the  new 
critical density  
\begin{equation}
\rho_{c}=\rho_{IN}\left(1-\frac{2b_{3}^{2}}{9b_{4}}\right) < \rho_{IN}
\label{eqrhoc}
\end{equation}
is shifted to lower density in comparison to equilibrium $\rho_{IN}$.
Eq. \ref{eqrhoc} gives	 the expression for new transition density as given in main text.
Hence using renormalised mean field theory we find a  jump $\Delta S$ at a lower density
as compared to equilibrium I-N transition density.



\begin{thebibliography}{10}
%
\bibitem{harada} Y. Harada, A. Nogushi, A. Kishino and T. Yanagida, Nature (London) {\bf 326}, 805 (1987);
M. Badoual, F. Jülicher and J. Prost, Proc. Natl. Acad. Sci. U.S.A. {\bf 99}, 6696 (2002).
%
\bibitem{nedelec} F. J. Nédélec, T. Surrey, A. C. Maggs and S. Leibler, Nature (London) {\bf 389}, 305 (1997).
%
\bibitem{rauch} E. Rauch, M. Millonas and D. Chialvo, Phys. Lett. A {\bf 207}, 185 (1995).
%
\bibitem{benjacob} E. Ben-Jacob, I. Cohen, O. Shochet, A. Tenenbaum, A. Czirók and T. Vicsek, Phys. Rev. Lett. {\bf 75}, 2899 (1995).

\bibitem{animalgroups} {\it Animal groups in Three Dimensions}, edited by J. K. Parrish and W. M. Hamner (Cambridge University Press, Cambridge, 1997).
%
\bibitem{helbing} D. Helbing, I. Farkas and T. Vicsek, Nature {\bf 407}, 487 (2000);
D. Helbing, I. J. Farkas, and T. Vicsek, Phys. Rev. Lett. {\bf 84}, 1240 (2000).
%
\bibitem{feder} T. Feder, Phys. Today {\bf 60}(10), 28 (2007); C. Feare, {\it The Starlings} (Oxford University Press, Oxford, 1984).
%
\bibitem{kuusela31} E. Kuusela, J. M. Lahtinen and T. Ala-Nissila, Phys. Rev. Lett. {\bf 90}, 094502 (2003).
%
\bibitem{hubbard} S. Hubbard, P. Babak, S. Sigurdsson and K. Magnusson, Ecol. Modell. {\bf 174}, 359 (2004).
%
\bibitem{vnarayan}V. Narayan, N. Menon and S. Ramaswamy, J. Stat. Mech. P01005 (2006); V. Narayan, S. Ramaswamy, and N. Menon, Science {\bf 317}, 105 (2007).
%
\bibitem{kudrolli} D.L. Blair, T. Neicu and A. Kudrolli, Phys. Rev. E {\bf67}, 031303 (2003).
%
\bibitem{sriramrmp} M. C. Marchetti, J. F. Joanny, S. Ramaswamy, T. B. Liverpool, J. Prost, M. Rao, and R. A. Simha, Rev. Mod. Phys. {\bf 85}, 1143 (2013).
%
\bibitem{tonertusr} J. Toner, Y. Tu, and S. Ramaswamy, Ann. Phys. (Amsterdam) {\bf 318}, 170 (2005).
%
\bibitem{rev} S. Ramaswamy, Annu. Rev. Condens. Matter Phys. {\bf 1}, 323 (2010)
.
%
\bibitem{kemkemer} R. Kemkemer, D. Kling, D. Kaufmann and H. Gruler, Eur. Phys. J. E {\bf 1}, 215, (2000)
%
\bibitem{serra}  X. Serra-Picamal, V. Conte, R. Vincent, E. Anon, D. T. Tambe, E. Bazellieres, J. P. Butler, J. J. Fredberg and X. Trepat, Nat. Phys. {\bf 8}, 628 (2012).
%
\bibitem{schaller}  V. Schaller, C. Weber, C. Semmrich, E. Frey and A. R. Bausch, Nature (London) {\bf 467}, 73 (2010).
%
\bibitem{surrey} T. Surrey, F. J. Nédélec, S. Leibler and E. Karsenti,  Science {\bf 292}, 1167 (2001).
%
\bibitem{sradititoner} S. Ramaswamy, R. A. Simha and J. Toner, Europhys. Lett. {\bf 62}, 196 (2003).
%
\bibitem{shradhanjop} E. Bertin, H. Chat\'{e}, F. Ginelli, S. Mishra, A. Peshkov and S. Ramaswamy, New J. of Phys. {\bf 15}, 085032 (2013).
%
\bibitem{chateprl2006} H. Chat\'{e}, F. Ginelli, and R. Montagne
Phys. Rev. Lett. {\bf 96}, 180602  (2006). 
%
\bibitem{shradhatrans2014} S. Mishra, S. Puri, and S. Ramaswamy, Phil. Trans. of the Royal Soc. A: Mathematical, Physical and Engineering Sciences {\bf 372}, 20130364 (2014).
%
\bibitem{aparnaredner} E. Putzig, G. S. Redner, A. Baskaran, A. Baskaran, arXiv:1506.03501 (2015).
%
\bibitem{shimanatcomm} Xia-qing Shi  and Yu-qiang Ma, Nat. Comm. {\bf 4}, 3013 (2013).
%
\bibitem{yeomans}S. P. Thampi, R. Golestanian and J. M. Yeomans,  Euro. Phys. Lett. {\bf 105}, 18001 (2014).
%
\bibitem{ncommam} S. Jordens, L. Isa, I. Usov and R. Mezzenga, Nat. Comm. {\bf 4}, 1917 (2013).
%
\bibitem{pgdgenne} P. G. de Gennes and J. Prost, {\it The Physics of Liquid Crystals} (Clarendon Press, Oxford, 1995).
%
\bibitem{llasher} A. Lebwohl and G. Lasher, Phys. Rev. A {\bf 6}, 426 (1972).
%
\bibitem{dilutedxymodel} Y. M. Blanter, Y. E. Lozovik and A. Y. Morozov, Phys. Scr. {\bf 52}, 237 (1995); S. A. Leonel, P. Z. Coura, A. R. Pereira, L. A. S. Mól, and B. V. Costa, Phys. Rev. B {\bf 67}, 104426 (2003).
%
\bibitem{mcbinder} D. P. Landau and K. Binder, {\it A Guide to Monte Carlo Simulations in Statistical Physics}, Second edition (Cambridge University Press, Cambridge, 2005).
%
\bibitem{SM}  Supplementary Material for (I) model figure, 
(II) estimate of critical temperature $T_c(C)$ 
in equilibrium model and (III) renormalised mean field 
calculation of effective free energy $f_{eff}(S)$ for small $S$. 
%
\bibitem{paxton} W. F. Paxton, K. C. Kistler, C. C. Olmeda, A. Sen, S. K. St. Angelo, Y. Cao, T. E. Mallouk, P. E. Lammert and V. H. Crespi, J. Am. Chem. Soc. {\bf 126}, 13424 (2004).
%
\bibitem{chate} H. Chat\'{e}, F. Ginelli, G.  Gr\'{e}goire and F. Raynaud, Phys. Rev. E {\bf 77}, 046113 (2008).
%
\bibitem{shradhapre} S. Mishra, A. Baskaran and M. C. Marchetti, Phys. Rev. E {\bf 81}, 061916 (2010).
%
\bibitem{plateauphase} M. Takigawa and F. Mila, {\it Introduction to Frustrated Magnetism: Materials, Experiments, Theory} (Springer, Heidelberg, 2011), ch 10
%
\bibitem{shimaprl2011} Xia-qing Shi, Yu-qiang Ma, arXiv:1011.5408 (2010).
%
\bibitem{coleman} S. Coleman and E. Weinberg, Phys. Rev. D {\bf 7}, 1888 (1973).
%
\bibitem{fdfopt} B. I. Halperin, T. C. Lubensky and S. K. Ma, Phys. Rev. Lett. {\bf 32}, 292 (1974); J. H. Chen, T. C. Lubensky and D. R. Nelson, Phys. Rev. B {\bf 17}, 4274 (1978).
%
\bibitem{Berezinskii} V. L. Berezinskii, Sov. Phys. JETP {\bf 34}(3), 610 (1972).
%
\bibitem{KT} M. Kosterlitz and D. Thouless, J. Phys. C {\bf 6}, 1181 (1973).
%
\bibitem{wulib} Xiao-Lun Wu and A. Libchaber, Phys. Rev. Lett., {\bf 84}, 3017 (2000).
%
%
\end{thebibliography}

\begin{thebibliography}{10}

\bibitem{pgdgenne} P. G. de Gennes and J. Prost, {\it The Physics of Liquid Crystals} (Clarendon Press, Oxford, 1995).
%
%
%
\bibitem{njopshradha} E. Bertin, H. Chat\'{e}, F. Ginelli, S. Mishra, A. Peshkov and S. Ramaswamy, New J. of Phys. {\bf 15}, 085032 (2013).
%
\bibitem{chaiklub} P. M. Chaikin and T. C. Lubensky, {\it Principles of Condensed Matter Physics} (Cambridge University Press, Cambridge, 2000).
%

%
\end{thebibliography}
\end{document}